\documentclass[11pt,twoside]{article}

\usepackage{asp2006}
\usepackage{graphicx}
\usepackage{natbib}

\begin{document}
\title{Large-scale Propagation of Very Light Jets in Galaxy Clusters} 
\author{V. Gaibler, M. Camenzind}
\affil{Landessternwarte, ZAH, K\"onigstuhl 12, 69117 Heidelberg, Germany}
\author{M. Krause}
\affil{Astrophysics Group, Cavendish Laboratory, Madingley Road,
      Cambridge CB3 0HE, United Kingdom}

\begin{abstract} 
  We performed MHD simulations of very light bipolar jets with density contrasts
  down to $10^{-4}$ in axisymmetry, which were injected into a medium of constant
  density and evolved up to $200$ kpc ($200\: r_{\mathrm{j}}$) full length.
  These jets show weak and roundish bow shocks as well as broad cocoons and
  thermalize their kinetic energy very efficiently. We argue that very light
  jets are necessary to match low-frequency radio observations of radio lobes as
  well as the bow shocks seen in X-rays. Due to the slow propagation, the
  backflows and their turbulent interaction in the midplane are important for a
  realistic global appearance.
\end{abstract}

\section{Introduction}

During the last years, simulations of extragalactic jets with reasonable
resolution and realistic sizes became computationally feasible, which makes
comparisons between simulated and observed properties possible
\citep{Saxton2002,Zanni2003,Carvalho2005,KrauseVLJ2,ONeill2005}. Unfortunately,
the direct physical variables and the observed properties are rather hard to
link, which leaves simulations with a wide range of parameters. Simulations are
mainly governed by the initial setup of the density ratio between jet and
ambient gas, the Mach number and the magnetic field. If the magnetic field is
not dynamically dominant (though important), the density contrast is the most
dominant parameter, but may be one of the hardest to measure. The thermal
jet pressure has turned out to be of little importance in the very light jet
limit \citep{KrauseVLJ1}. As the (kinetic) power of a jet can be estimated from
energies in X-ray bubbles, typical values of velocity, lifetime, jet radius and
cluster gas densities indicate that density contrasts of $10^{-2}$ to $10^{-4}$
(or even lower) are necessary to describe real sources. Parameter studies
support this further, if the global jet/cocoon/bow shock properties are
compared. Thus, we concentrate on the very light jets with magnetic fields as
another important ingredient.

\section{Numerical Method and Setup}

We examine the evolution of the jets in axisymmetric simulations using the
nonrelativistic MHD code NIRVANA \citep{ZY97} and evolve the magnetic
fields using the constrained transport method, which conserves
$\nabla\cdot\mathbf{B}$ to machine roundoff errors.  

Simulations of very light hydro and MHD jets were performed with
density contrasts $\eta = \rho_\mathrm{j}/\rho_\mathrm{a}$ between $10^{-4}$ and
$10^{-1}$ where the jet density is $\rho_\mathrm{j}$ and the ambient gas has a
constant density of $\rho_\mathrm{a}$. We will focus 
on the MHD jets, as their hydro counterparts are only for comparison. 
The bipolar jet was injected along the Z axis in cylindrical coordinates with a
jet radius of $r_\mathrm{j}=1$ kpc, the jet speed and the sound speed were
fixed at $0.6$ c and $0.1$ c respectively (which gives internal Mach number 6). 
The ambient gas has a density of
$0.01\:\mathrm{m_p/cm^{3}}$ and a temperature of $5\times 10^7$ K. Fully
ionized hydrogen was assumed for both the jet and the external medium. The MHD
simulations have an initial dipolar field in the whole domain with $\sim
20\:\mu\mathrm{G}$ at the jet boundary and a temporally constant toroidal field
with $\sim 15\:\mu\mathrm{G}$ which is confined to the nozzle. For the $10^{-3}$
and especially the $10^{-4}$ jet, the magnetic fields thus become dynamically
important and influence the appearance. The simulations were run until they
reach the boundary of the grid which has $(4000 \times 800)$ or $(4000 \times
1600)$ cells (depending on the density contrast) and the jet radius is resolved
with 20 cells.

\section{Morphology}

\begin{figure}[h]
  \centering
  \includegraphics[width=0.8\textwidth]{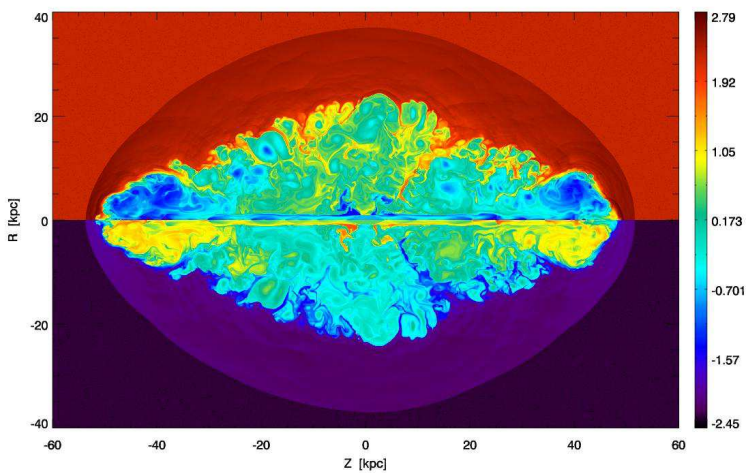}
  \caption{Density and temperature for a bipolar jet with $\eta=10^{-3}$ after
  15 Myrs.  The upper panel shows the density in units of $10^{-28}$ g/cm$^3$,
  the lower one shows the logarithm of temperature in units of $10^{10}$ K.}
  \label{fig:dentemp1e-3}
\end{figure}

Density and temperature distribution for a $\eta=10^{-3}$ jet is shown in
Fig.~\ref{fig:dentemp1e-3}. The jet backflow blows up a pronounced cocoon,
surrounded by a thick shell of shocked ambient matter. Ambient gas is mixed into
the cocoon in finger-like structures due to Kelvin-Helmholtz instabilities at
the contact surface. Near the jet heads, this instability is suppressed by the
magnetic field, which leads to a smoother appearance there. In purely
hydrodynamic simulations, this stabilization is absent. As observations at low
radio frequencies show quite smooth contact discontinuities, this indicates the
importance of magnetic fields there.

The cocoon is highly turbulent and vortices hitting the jet beam can easily
destabilize and disrupt it for low jet densities. The Mach numbers quickly
decrease and there is no classical ``Mach disk'' anymore -- the terminal shock
moves back and forth and isn't well-defined.

Because very light jets only propagate slowly, the backflow is strong and the
turbulence makes the interaction between both jets in the midplane important.
These jets have to be simulated bipolarly to get the lateral expansion and hence
the global appearance right. If only one jet was simulated, the result would 
strongly depend on the boundary condition in the equatorial plane 
\citep[as shown in][]{Saxton2002}.

Outside of the contact surface is the shocked ambient gas, which is pushed
outwards by the cocoon pressure. The bow shock for very light jets is different
in its shape and strength from that of heavier jets (see
section~\ref{sec:bowcoc}). It is additionally changed by a density profile in
the external medium \citep{KrauseVLJ2}, which increases the aspect ratio with time
and shows cylindrical cocoons.

\begin{figure}[tb]
  \centering
  \includegraphics[width=0.8\textwidth]{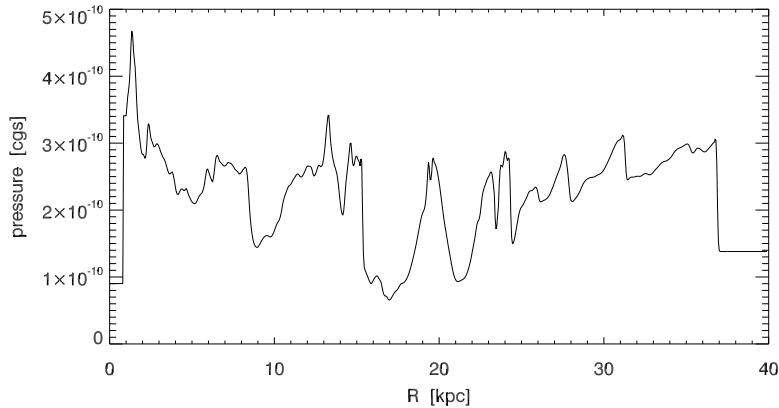}
  \caption{Pressure slice through the $\eta=10^{-3}$ jet at $Z=0$ after 15 Myr
  (as Fig.~\ref{fig:etacomp}). The bow shock is located at $R \approx
  37$ kpc.}
  \label{fig:pressureslice}
\end{figure}
As example, a radial pressure slice at $Z=0$ is shown for the $10^{-3}$ jet
(Fig.~\ref{fig:pressureslice}). The pressure jump at $R \approx 37$ kpc is the
bow shock and is pretty weak compared to bow shocks in heavier jets. Shock
speed, pressure and density jump, consistently with the shock jump conditions,
give a Mach number of $1.4$. 

Observations of bow shocks \citep[e.g. Hercules A in][]{Nulsen2005}, which are
possible with modern X-ray telescopes, show low Mach
numbers and low ellipticity, thus supporting the necessity for very light jet
parameters. To find the right cocoon shapes, for comparison low-frequency radio
observations have to be chosen, because at higher frequencies only a small part
of the cocoon is visible as radio lobes (cooled-down electron population in the
backflow is invisible at these frequencies).

\section{Pressure Evolution}

\begin{figure}
  \centering
  \includegraphics[width=0.8\textwidth]{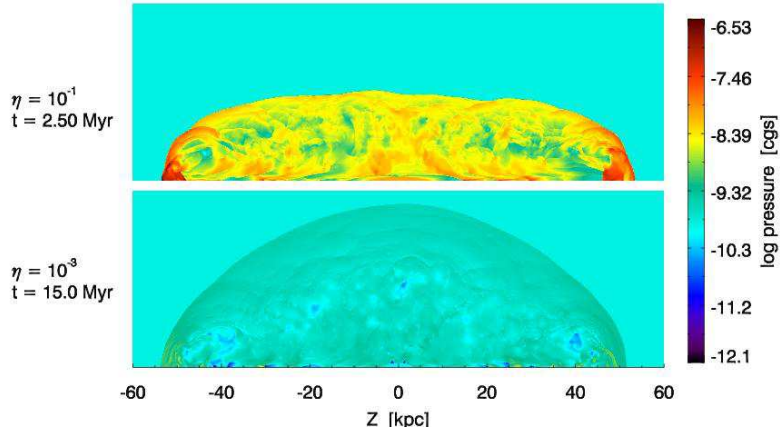}
  \caption{Pressure maps for a $\eta=10^{-1}$ and a $10^{-3}$ jet with the same lengths.
  The pressure is shown logarithmically in dyne/cm$^2$. The heavier jet
  is still much overpressured with respect to the ambient gas and has a much
  more elongated bow shock compared to the elliptically-shaped bow shock for the
  lighter jet.}
  \label{fig:etacomp}
\end{figure}
The pressure slice shows many variations in Fig.~\ref{fig:pressureslice}, which
is not surprising if one looks at the turbulent motion and the mixing inside
the cocoon in Fig.~\ref{fig:dentemp1e-3}. Strong pressure waves travel
through the cocoon and try to find pressure balance. This process is much more
effective for very light jets due to the much slower jet head propagation
and it leads to a rather spherical expansion of the bow shock, just like an overpressured
bubble. The cocoon of the $10^{-1}$ jet in Fig.~\ref{fig:etacomp} is
overpressured by a factor of $20$ with respect to the ambient gas, while being
a factor of only $1.5$ for the $10^{-3}$ jet (and $4.9$ for this jet at
$t = 2.5$ Myr).
\begin{figure}[t]
  \centering
  \includegraphics[angle=0,width=0.45\textwidth]{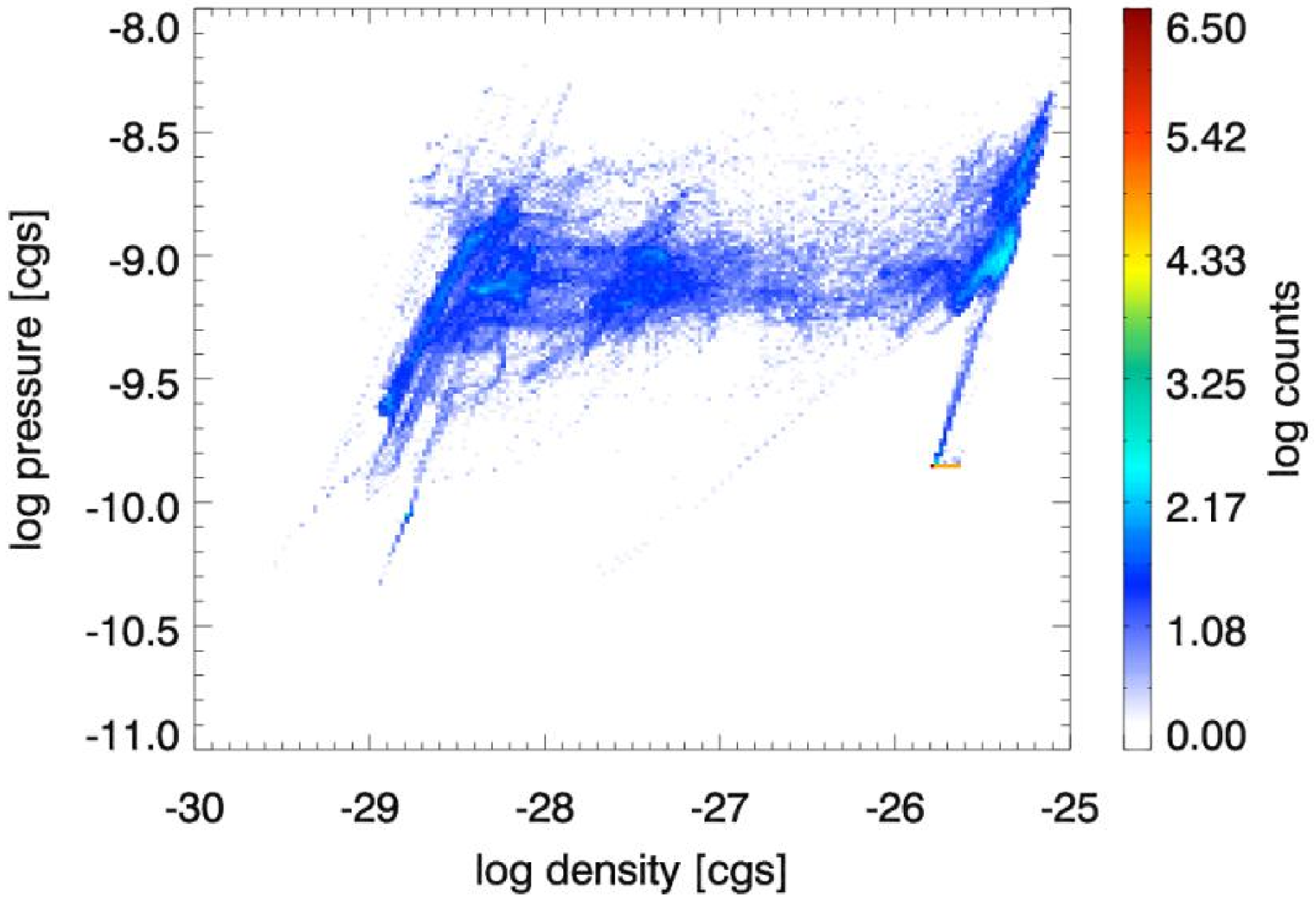}
  \includegraphics[angle=0,width=0.45\textwidth]{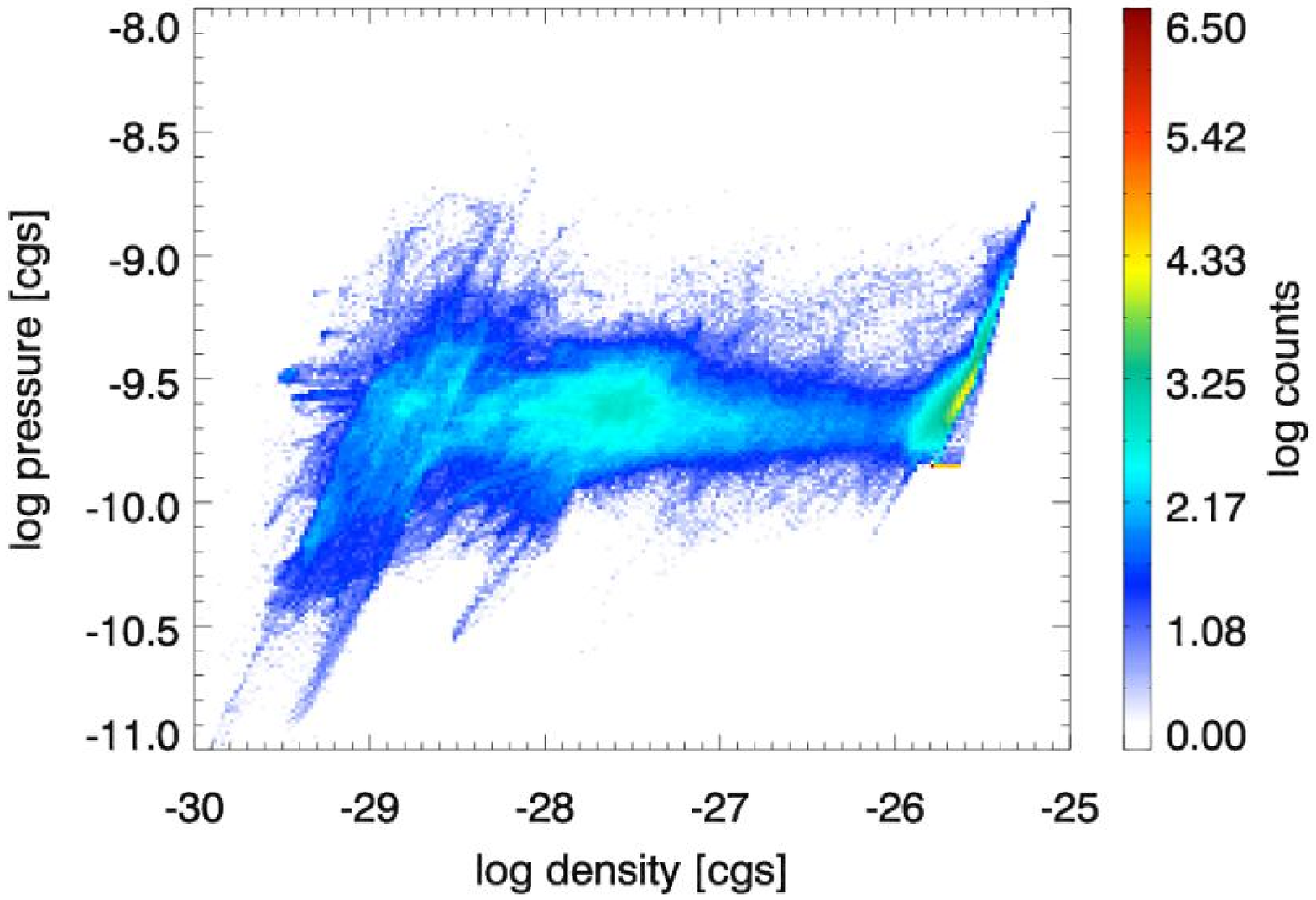}
  \caption{Pressure vs.\ density histogram for a $10^{-3}$ jet after
  $2$ Myr (left panel) and $15$ Myr (right panel). Cell counts, pressure and
  density are shown logarithmically using cgs units.}
  \label{fig:pndiag}
\end{figure}

This can also be seen in the pressure--density diagrams in Fig.~\ref{fig:pndiag}.
The ambient gas is described by the patch near $(-26, -10)$, the jet nozzle by
the cells around $(-29,-10$). Adiabatic compression and expansion leads to
the oblique and longish features present at different positions. Top right of
the jet nozzle position are the cocoon grid points, which spread over a large
range of density to the right because of mixing with shocked ambient gas, which
is the elongated feature top right of the ambient gas position. Comparing the two
different simulation snapshots, we find that the pressure distribution is
quickly adjusting towards the external pressure, in agreement to the findings in
\cite{KrauseVLJ1}.

\section{Bow Shock and Cocoon}
\label{sec:bowcoc}

\begin{figure}
  \centering
  \includegraphics[width=0.8\textwidth]{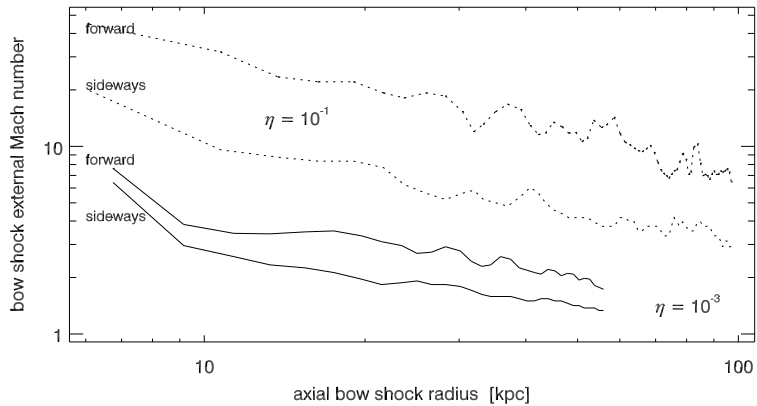}
  \caption{Evolution of the forward and sideways bow shock strength for density
  contrasts $\eta = 10^{-1}$ (dotted) and $10^{-3}$ (solid) as a function of the
  monotonically increasing axial bow shock radius.}
  \label{fig:bsstrengthevol}
\end{figure}

The quick decrease in cocoon pressure naturally affects the strength of the bow
shock as it is this pressure that drives the shock sideways.
Fig.~\ref{fig:bsstrengthevol} shows the temporal evolution of the bow shock
strength, in terms of external Mach numbers, for the forward ($R=0$) direction
as well as the sideways ($Z=0$) direction for jets with different density
contrasts. For easier comparison with observations, the axial bow shock radius
is used for the abscissa instead of time (but both increase monotonically).

The bow shocks in forward direction are always stronger than the
sideways shocks due to the direct impact of the jet onto the ambient gas. The
lighter jet has a much weaker bow shock in all directions and the differences
between the two directions shown are much less pronounced.
The axial diameter of the bow shock increases proportionally to $t^{0.68}$ after
a slower growth during the first $2$ Myr, the width grows similarly as
$t^{0.64}$. An exponent of 0.6 is expected for the blast wave expansion with
constant jet power \citep{KrauseVLJ1}, while the slower growth rate in the
initial phase behaves more like a Sedov blast wave (fixed initial energy amount,
exponent is 0.4).
\begin{figure}
  \centering
  \includegraphics[width=0.8\textwidth]{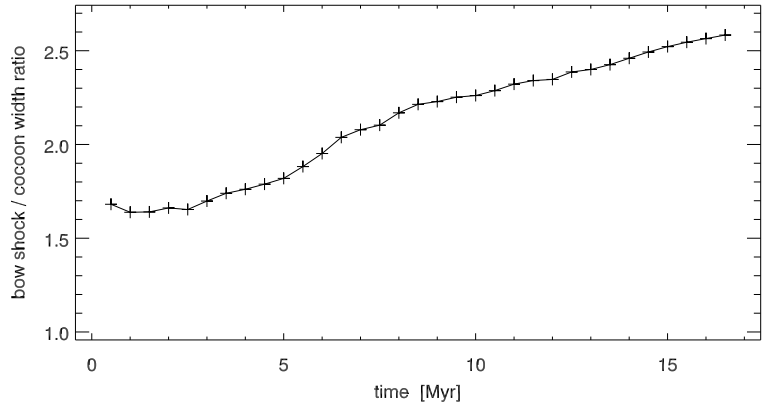}
  \caption{Bow shock / cocoon width ratio over time for a $10^{-3}$ jet.}
  \label{fig:bscowidths}
\end{figure}

The cocoons, measured by their full (bipolar) lengths and their full Z-averaged
widths, grow like the bow shock in axial direction ($\propto t^{0.71}$), but
much slower in width ($\propto t^{0.38}$). This leads to a continuously
increasing bow shock vs.\ cocoon width ratio (Fig.~\ref{fig:bscowidths}) with a
very thick layer of shocked ambient gas. This effect is weak for heavier jets,
but more proncounced the lighter the jet is.

\begin{figure}
  \centering
  \includegraphics[width=0.8\textwidth]{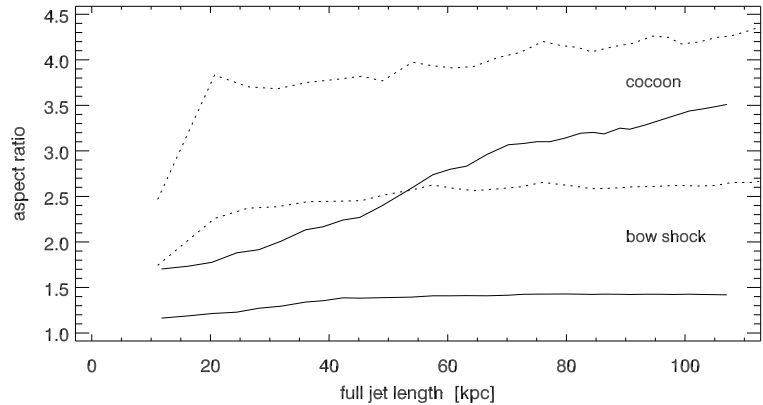}
  \caption{Aspect ratios over full jet length. Dotted line: $10^{-1}$ jet,
  solid: $10^{-3}$ jet. In each case, the lower lines refer to the bow shock and
  the upper ones to the cocoon.}
  \label{fig:aspectoverlength}
\end{figure}

The aspect ratios (length/width) for the $10^{-1}$ and $10^{-3}$ jets are
plotted in Fig.~\ref{fig:aspectoverlength}. The bow shock for the lighter jet
starts with a roughly spherical shape and only slightly increases its aspect
ratio (length/width) to a constant value of $1.4$
(Fig.~\ref{fig:aspectoverlength}). The heavier jet behaves similarly but
approaches a much higher aspect ratio of 2.6. Thus the aspect ratio of the bow
shock may be a good property to compare with observations. The aspect ratio of the
cocoons in contrast continues to increase, with the heavier jet being on much
higher values at all times.

\section{Thermalization}

From the quick adjustment of pressure towards an average value, one might expect
a strong conversion of the (kinetic) jet power to thermal energy. This, in fact,
is measured for our simulations. Already for the heavy $10^{-1}$ jet, on average
$67\,\%$ of the total energy input is measured as thermal energy increase and
only $33\,\%$ as kinetic energy increase. For the $10^{-2}$ jet this is $72\,\%$
vs.\ $27\,\%$, and for the $10^{-3}$ jet already $81\,\%$ of the jet power
appear as thermal energy increase with $16\,\%$ going into kinetic energy. Here,
already $2\,\%$ go into an increase in magnetic energy, because with lower
density the magnetic fields of constant values become more and more important.
The $10^{-4}$ simulation showed a thermalization efficiency of $94\,\%$, but the
fractions for kinetic and magnetic energy are now governed by the strong
magnetic pressure and will be examined in the future. 

The overall trend to very efficient thermalization for low-density jets nicely
suits the increasingly spherical bow shock shape due to (isotropic) cocoon
pressure. It also provides the cluster with a huge amount of thermal energy and
high-entropy plasma, which may be relevant for the problem of
cluster heating and cooling flows \citep[eg.][]{Magliocchetti2007}.

\acknowledgements 
This work was also supported by the Deutsche For\-schungs\-ge\-mein\-schaft
(Sonderforschungsbereich 439).


\begin{thebibliography}{}
  \bibitem[Carvalho et al.(2005)]{Carvalho2005}
    Carvalho, J.~C., Daly, 
    R.~A., Mory, M.~P., \& O'Dea, C.~P.\ 2005, \apj, 620, 126
  \bibitem[Krause(2003)]{KrauseVLJ1}
    Krause, M.\ 2003, \aap, 398, 113
  \bibitem[Krause(2005)]{KrauseVLJ2}
    Krause, M.\ 2005, \aap, 431, 45
  \bibitem[Magliocchetti \& Br{\"u}ggen(2007)]{Magliocchetti2007}
    Magliocchetti, M., \& Br{\"u}ggen, M.\ 2007, \mnras, 528
  \bibitem[Nulsen et al.(2005)]{Nulsen2005}
    Nulsen, P. E. J., Hambrick, D. C., McNamara, B. R., Rafferty, D.,
    Birzan, L., Wise, M. W., \& David, L. P. 2005, \apj, 625, L9
  \bibitem[O'Neill et al.(2005)]{ONeill2005} O'Neill, S.~M., Tregillis, I.~L.,
    Jones, T.~W., \& Ryu, D.\ 2005, \apj, 633, 717
  \bibitem[Saxton et al.(2002)]{Saxton2002} Saxton, C.~J., 
    Sutherland, R.~S., Bicknell, G.~V., Blanchet, G.~F., \& Wagner, S.~J.\ 
    2002, \aap, 393, 765
  \bibitem[Zanni et al.(2003)]{Zanni2003} Zanni, C., Bodo, G., 
    Rossi, P., Massaglia, S., Durbala, A., \& Ferrari, A.\ 2003, \aap, 402, 949
  \bibitem[Ziegler \& Yorke(1997)]{ZY97}
    Ziegler, U., \& Yorke, H.~W. 1997,
    Computer Physics Communications, 101, 54
\end{thebibliography}
\end{document}